\documentclass[%
 reprint,
 superscriptaddress,
 amsmath,amssymb,
 aps,
]{revtex4-1}

\usepackage{graphicx}
\usepackage{dcolumn}
\usepackage{bm}
\usepackage[utf8]{inputenc}
\begin{document}


\title{3D superconducting gap in FeSe from ARPES}

\author{Y. S. Kushnirenko}
 \affiliation{IFW Dresden, Helmholtzstr. 20, 01069 Dresden, Germany}
\author{A. V. Fedorov}
 \affiliation{IFW Dresden, Helmholtzstr. 20, 01069 Dresden, Germany}
\author{E. Haubold}
 \affiliation{IFW Dresden, Helmholtzstr. 20, 01069 Dresden, Germany}
\author{S. Thirupathaiah}
 \affiliation{IFW Dresden, Helmholtzstr. 20, 01069 Dresden, Germany}
 \affiliation{Solid State and Structural Chemistry Unit, Indian Institute of Science, Bangalore, Karnataka, 560012, India.}
\author{T. Wolf}
 \affiliation{Institute for Solid State Physics, Karlsruhe Institute of Technology, 76131 Karlsruhe, Germany}
\author{S. Aswartham}
 \affiliation{IFW Dresden, Helmholtzstr. 20, 01069 Dresden, Germany}
\author{I. Morozov}
\affiliation{IFW Dresden, Helmholtzstr. 20, 01069 Dresden, Germany}
 \affiliation{Lomonosov Moscow State University, 119991 Moscow, Russia}
\author{T. K. Kim}
 \affiliation{Diamond Light Source, Harwell Campus, Didcot OX11 0DE, United Kingdom.}
\author{B. Büchner}
 \affiliation{IFW Dresden, Helmholtzstr. 20, 01069 Dresden, Germany}
\author{S. V. Borisenko}
 \affiliation{IFW Dresden, Helmholtzstr. 20, 01069 Dresden, Germany}


\begin{abstract}
We present a systematic angle-resolved photoemission spectroscopy study of the superconducting gap in FeSe. The gap function is determined in a full Brillouin zone including all Fermi surfaces and k$_z$-dependence. We find significant anisotropy of the superconducting gap in all momentum directions. While the in-plane anisotropy can be explained by both, nematicity-induced pairing anisotropy and orbital-selective pairing, the k$_z$-anisotropy requires additional refinement of theoretical approaches.
\end{abstract}

\pacs{Valid PACS appear here}
\maketitle

\medskip

Detailed knowledge of the gap function in iron-based superconductors can help to identify the mechanism of superconductivity in these materials. Recent clarification of the details of the electronic structure in FeSe to a precision of 1 meV \cite{fedorov16, Borisenko16NPh, watson2016evidence, pustovit16LTP, watson15PRB, zhang15PRB, suzuki2015momentum, ye15arXiv, nakayama14PRL, shimojima14PRB, maletz14PRB} is a necessary prerequisite to study the superconducting gap by angle-resolved photoemission spectroscopy (ARPES) and makes this material a perfect candidate for such a detailed investigation. There are several experimental studies of the superconducting gap in FeSe and closely related compounds using different  techniques: tunneling spectroscopy \cite{jiao2017superconducting, sprau2017discovery, moore2015evolution, kasahara2014field, singh2013spatial, song2011direct}, ARPES \cite{xu2016highly, hashimoto2018superconducting, okazaki2012evidence,miao2012isotropic}, specific heat \cite{sun2017symmetry,zeng2010anisotropic,abdel2015superconducting} and London penetration depth \cite{kasahara2014field,abdel2015superconducting}; but the consensus as for the size and symmetry of the gap in the full Brillouin Zone (BZ) has not been reached. Although ARPES still remains the only, though non-phase-sensitive, method of direct determination of the gap as a function of momentum, i.e. the gap function, the agreement between existing reports is far from being perfect. For instance, the authors of Ref. \onlinecite{miao2012isotropic} reported an isotropic 2.5 meV gap on a central pocket in FeSe$_{0,45}$Te$_{0.55}$, while the smaller  and considerably more anisotropic gaps in the compound with very similar composition FeSe$_{0.4}$Te$_{0.6}$ are found in Ref. \onlinecite{okazaki2012evidence}. As for the pristine FeSe, according to Ref. \onlinecite{hashimoto2018superconducting} the gap is anisotropic on a central hole-like pocket, as well as in slightly S- doped FeSe \cite{xu2016highly}, while no gap has been observed on the electron pockets in the corner of the BZ. This is in contrast to a majority of the tunneling results \cite{jiao2017superconducting,sprau2017discovery,moore2015evolution,kasahara2014field}, which imply a presence of multiple superconducting gaps. Moreover,  specific heat and London penetration depth also indicate the presence of two gaps \cite{abdel2015superconducting}. Finally, no study managed to shed any light on a possible k$_z$-dependence of the gap function, although it was mentioned in Ref.\onlinecite{xu2016highly} that no gap could be detected neither on the electron-like Fermi surfaces nor on part of the hole-like pocket near \(\Gamma\)-point. Therefore it is important to have a precise information about the behavior of the gap function throughout the whole BZ, preferably obtained by the same technique.

In this Letter, we report the results of a high-resolution ARPES study of the superconducting gap function in the single-crystals of FeSe, exactly from the same material in which we recently clarified the fine details of the electronic structure \cite{kushnirenko2017anomalous, fedorov16, Borisenko16NPh}. We clearly observed two anisotropic SC gaps on hole- and electron-like Fermi surfaces. Their momentum variation as a function of k$_x$, k$_y$ and k$_z$ is compared to the ones predicted by  orbital-selective pairing \cite{sprau2017discovery} and nematicity-induced anisotropy of the pairing gap \cite{Kang_arxiv}.

ARPES data have been collected at I05 beamline of Diamond Light Source \cite{hoesch17RSI}. Single-crystal samples were cleaved \textit{in situ} in a vacuum better than $2\times10^{-10}$ mbar and measured at temperatures ranging from 5.7 K. Measurements were performed using linearly polarized synchrotron light, utilizing Scienta R4000 hemispherical electron energy analyzer with an angular resolution of 0.2$^\circ$ – 0.5$^\circ$ and an energy resolution of 3 meV. Samples were grown by the KCl/AlCl$_3$ chemical vapor transport method .

\begin{figure*}[t]
	\centering
    \includegraphics[width=0.9\linewidth]{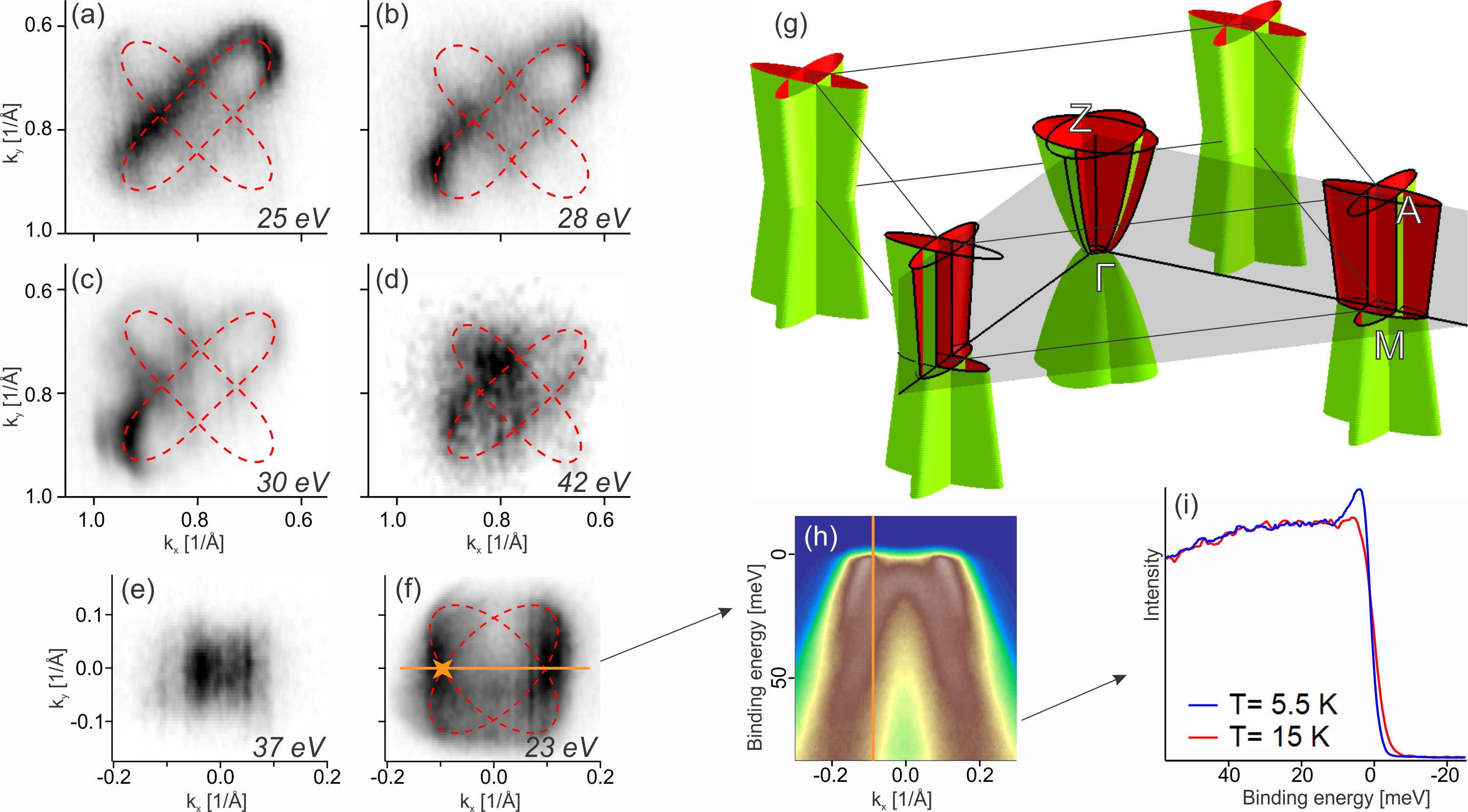}
	\caption{ (a)-(d) Fermi surface maps of the electron-like pockets measured using different photon energies. (e), (f) Fermi surface maps of the hole-like pockets measured using different photon energies. (g) Schematic sketch of the experimentally determined 3D Fermi surface of FeSe. (h) Momentum-energy intensity distribution along the line indicated in (f), (i) k$_F$ energy distribution curves above and below T$_c$ corresponding to the line from (h) and a star from (f).}
	\label{fig:one}
\end{figure*}

In Fig.\ref{fig:one}(a-f)  we show the experimental Fermi surface maps of electron- and hole-like pockets measured with different photon energies which correspond to different k$_z$-values in 3D BZ. A schematic picture of 3D Fermi surface of FeSe summarizing these and previous ARPES results is presented in Fig. \ref{fig:one}(g) and the evidence for the sensitivity of our experiments to the superconductivity itself is given in panels Fig.\ref{fig:one}(h,i). 

In the center of the BZ near Z-point, we see two hole-like elliptical pockets crossing each other (Fig. \ref{fig:one}(f)). These two  ellipses originate from two different domain orientations in the nematic state \cite{fedorov16,watson2016evidence,xu2016highly,pustovit16LTP,suzuki2015momentum,watson15PRB} because the surface area probed by photons is bigger than a typical domain size (above the nematic transition the FS near Z-point is a single rounded hole-like pocket). As one approaches the \(\Gamma\)-point, the size of these squeezed by nematicity hole-like pockets rapidly decreases resulting in a very small FS at k$_z$=0, as shown in Fig. \ref{fig:one}(e). Intensity distribution along the cut through the FS centered at Z-point clearly shows two sets of spin-orbit split $d_{xz,yz}$ dispersing features in Fig. \ref{fig:one}(h) while $d_{xy}$-band, which tops in FeSe at 50 meV below Fermi level, is hardly visible making itself noticeable only via hybridization with other states. The two features dispersing towards the Fermi level do not actually cross it and demonstrate all typical signs of the opening of the small superconducting gap. The direct evidence is given in Fig. \ref{fig:one}(i) by two energy distribution curves (EDC) taken above and below the critical temperature of FeSe. Emerging of a coherence peak as well as a typical shift of the leading edge midpoint are clearly observed. Because of the closely separated multiple features with drastically different Fermi velocities close to the Fermi level, it is the shift of the leading edge, or leading edge gap (LEG), which we will use throughout this paper to characterize the superconducting gap in FeSe.

In the corner of the BZ, there are two peanut-like \footnotemark[100] pockets crossing each other [Figs. \ref{fig:one}(a-d, g)]. The size of these electron-like Fermi surfaces is also changing with $k_z$. Going from A-point to M-point, one notices a significant shrinking. A popular interpretation \cite{watson15PRB,zhang15PRB,shimojima14PRB,suzuki2015momentum} of the presence of these two pockets is that they are the result of superposition of single electron pockets from different domains of the twinned sample, as is the case with the hole-pockets in the center (see above). In this approach, the other electron pocket, though expected in the conventional band structure calculations, should disappear below the transition. Our interpretation is that both pockets remain present also in the nematic phase and also in the single-domain sample, i.e. in agreement with the band structure calculations. The overlapping of the contributions from two different domains does take place as well, but resulting FS is more complicated, being a superposition of two sets of crossed "peanuts". Since each set becomes C$_2$-symmetric in the orthorhombic (nematic) phase, the overlap of such two structures rotated with respect to each other by $\sim$ 90$^\circ$ leads to an apparent doubling of each pocket. Such a small difference between Fermi surfaces is predicted by DFT calculations for the orthorhombic state (see Figs. 2(b-d) in Ref. \onlinecite{fedorov16}) and is in agreement with our previous ARPES results (see Figs. 2(e-g) in Ref. \onlinecite{fedorov16}) where the elusive doubling can still be resolved. In the Supplemental Material section \footnotemark[100] we present more ARPES data which unambiguously prove the presence of two sets of crossed peanuts Fermi surfaces in FeSe. In order to avoid additional complications with the extraction of gap values from the spectra, we adjust the photon energy and geometry of the experiment such that only one set of pockets is visible at a time.

\begin{figure}[b]
	\centering
    \includegraphics[width=1\linewidth]{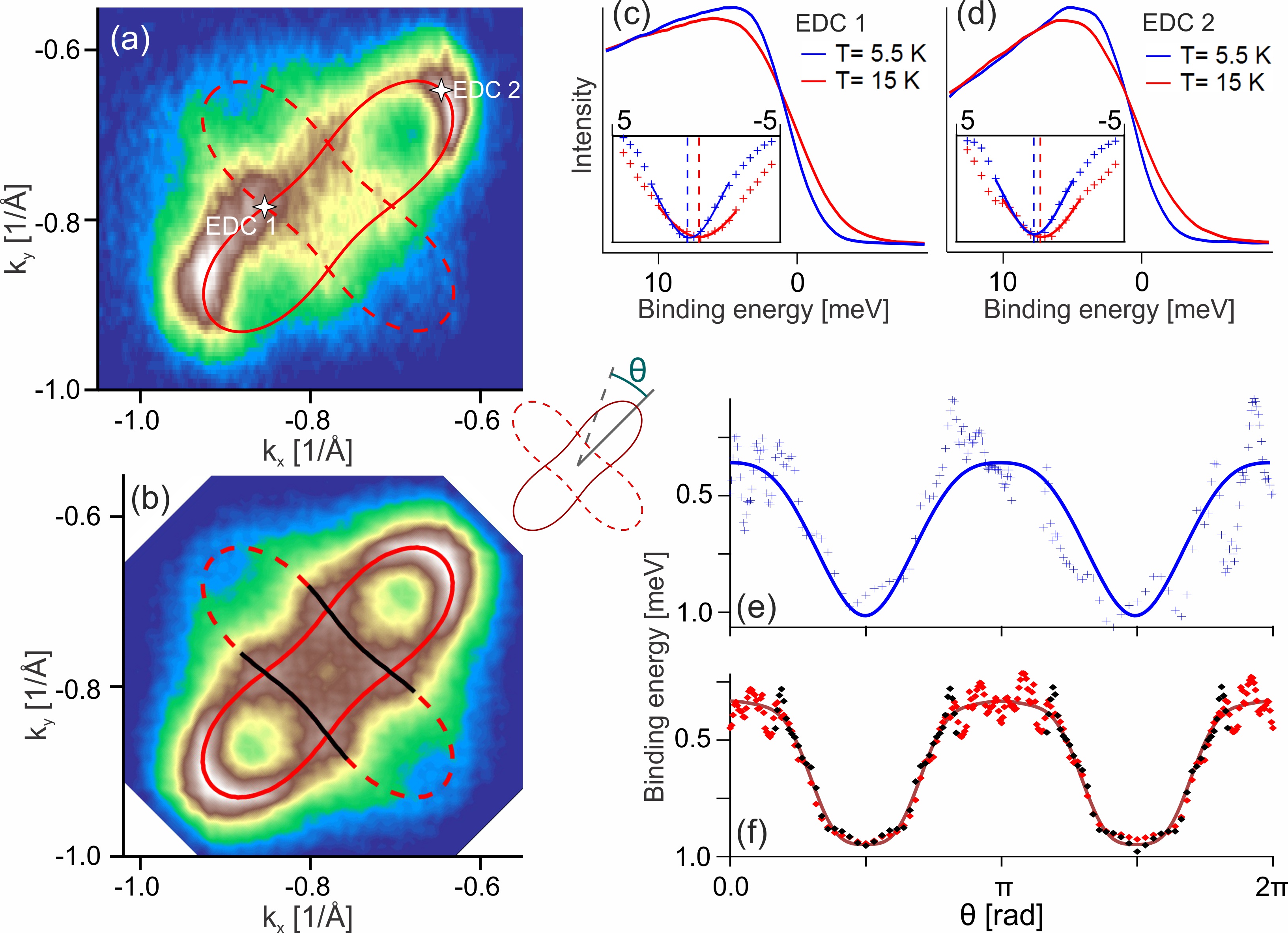}
	\caption{ (a) Fermi surface map of the electron-like pocket near A-point. (b) Symmetrized version of a map from (a). (c),(d) k$_F$ energy distribution curves from parts of the pocket marked by stars in (a) measured in the normal and superconducting states. Insets show the first derivatives of the same curves. (e) binding energy of the leading edge of k$_F$-EDCs on the electron-like pocket from row data. (f) The same as (e) but from the symmetrized dataset. In order to match results from two different ellipses, black dots are shifted by 90 degrees.}
	\label{fig:two}
\end{figure}

Now let us turn to the momentum variation of the superconducting gap on electron-like pocket near A-point. The data-set shown in Fig. \ref{fig:two}(a) was measured using 28 eV photons \footnotemark[101] with linear horizontal polarization from the sample cooled down to 5.5 K, i.e. in its superconducting state. We took the scans in a direction perpendicular to the BZ borders deliberately. Such experimental geometry allowed us to detect more spectral weight and thus more details in comparison with the previous ARPES studies of the SC gap in FeSe. Specifically, the photoemission intensity from the pockets' ends is not suppressed as in earlier studies and even some parts of the other pocket are visible. First, we establish the presence of the superconducting gap also in this region of momentum space. Figs. \ref{fig:two}(c,d) show k$_F$-EDCs from places on the pocket marked with stars on Fig. \ref{fig:two}(a) in the normal and the superconducting state. In both cases, the pairs of EDCs demonstrate the shift of the leading edge position to higher binding energies upon entering the superconducting state. The insets show the derivatives which peak at different binding energies, by definition representing the  positions of the leading edges. Moreover, this shift is not the same, it is equal to 0.3 meV and 0.6 meV respectively, which indicates that the gap not only is present on electron pockets, but also it is not isotropic. 

For further analysis of gap anisotropy we have extracted binding energy of the leading edge along the most intense peanut pocket. The result is shown in Fig. \ref{fig:two}(e). Already these result obtained from the raw data clearly demonstrate the presence of noticeable gap anisotropy. To compensate for the matrix element effects, which result in uneven intensity distribution along the peanut, we symmetrized the dataset from panel (a) with respect to the long axis of the better visible pocket. Symmetrized FS map is shown in Fig. \ref{fig:two}(b) and corresponding binding energy of the leading edge as a function of $\theta$ is shown in Fig. \ref{fig:two}(f). Here the red dots correspond to an ellipse from the map with a red contour on top, while the black dots correspond to a visible part of another peanut (black lines in the map). This plot presents direct evidence of the anisotropic superconducting gap on electron-like pocket near A-point of FeSe. Not surprising, the symmetry of this gap function is C$_2$. The largest gap, which corresponds to the lowest leading edge position, is on the shorter axis of the peanuts while the gap minima are on the longer axis of the peanuts. Fitting the data with a periodic function $\epsilon=A_0+A_1 cos(\phi)+A_2 cos(2\phi)$, where $A_0, A_1, A_2$ are free parameters, gives the amplitude of the gap variation of 0.6 meV (see brown curve in Fig. \ref{fig:two}(f)). The fit to the raw data from the non-symmetrized map has almost the same shape and amplitude (blue curve in Fig. \ref{fig:two}(e)).

As mentioned above, each peanut of electron-like FS is a superposition of 2 components which originate from two orthorhombic domains (for details see Supplemental Material \footnotemark[100]). It is thus instructive to know, which exactly component is analyzed in Fig. \ref{fig:two}. From the comparison of the pocket shape obtained from Fig. \ref{fig:two}(a) with the one in Ref. \footnotemark[100] one can conclude that intensity on this map mostly originates from the shorter peanut and we thus have analyzed the gap anisotropy related to this pocket. The finite intensity from the longer peanut, which is still present on the map, could effectively lower the leading-edge energy position at the ends (longer axis) of the shorter peanut \cite{BorisenkoSymm}. This is because the dispersions along this direction in the momentum space run nearly parallel to each other and the one which supports the longer peanut (secondary signal) can contribute spectral weight to k$_F$-EDC of the primary feature. Consequently, the amplitude of the LEG gap anisotropy can be underestimated. Here we would like to point out, that although LEG is a good qualitative measure of the superconducting gap and its anisotropy, the correspondence of the absolute values is more complicated and depends on many factors \cite{kordyuk2003measuring}. Since modelling of the spectral function, definitely necessary to provide (model-dependent) absolute values of the gap in the case of FeSe, is beyond the scope of this paper, we will continue to discuss LEG as a robust quantity, which can be extracted directly from the raw data without a sophisticated data analysis. As a rule, the real gap is slightly larger than LEG.  

\begin{table}[h]
    \centering
    \begin{tabular}{c|c}
        Photon energy & Leading-edge gap anisotropy\\ \hline
        25 eV & 0.35 meV \\ 
        28 eV  (A-point) & 0.6 meV \\
        30 eV & 0.8 meV \\ 
        42 eV  (M-point) & 0.7 meV \\
    \end{tabular}
    \caption{Leading-edge gap anisotropy on the electron-like pocket for different k$_z$ values}
    \label{tab:one}
\end{table}

To explore the gap function in whole 3D momentum space, we have also analyzed the datasets taken using different photon energies: 42 eV which corresponds to M-point \footnotemark[101], 25 eV and 30 eV. The amplitude of the LEG variation, i.e. the difference in the leading edge position between long and short peanut axis k$_F$-EDCs, is given in Table \ref{tab:one}. While the functional form of the anisotropy is approximately the same, there are clear oscillations of the amplitude as a function of k$_z$ with the most rapid variations taking place in the vicinity of the A-point.

\begin{figure}[]
	\centering
    \includegraphics[width=1\linewidth]{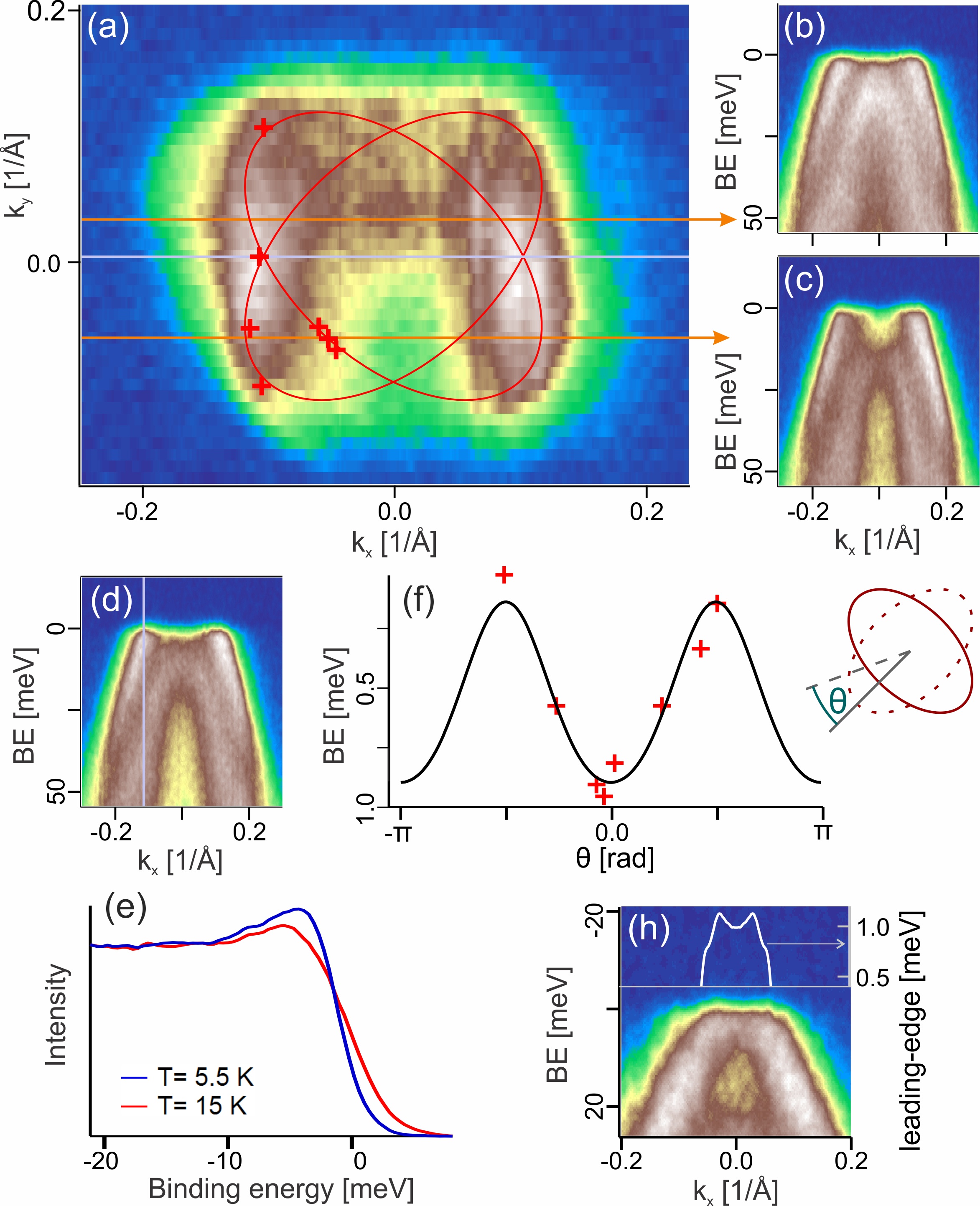}
	\caption{ (a) Fermi surface map of hole-like pocket near Z-point. (b)-(d) spectra measured in directions shown on the map with two orange and one grey lines respectively. (e) k$_F$ energy distribution curves obtained in a direction shown with a line on (c) from spectra measured in the normal and superconducting state. (f) binding energy of the leading edge of the k$_F$-EDCs from the hole-like pocket. (h) cut through the $\Gamma$-point with white line representing the leading edge position of EDCs near zero momentum.}
	\label{fig:three}
\end{figure}

Now let us consider the hole-like Fermi surface in the center of the Brillouin zone. Fig.\ref{fig:three}(a) shows Fermi surface map of the hole-like pocket near Z-point. Here it is convenient to analyze the EDCs from the lower part of the map, as panels (b) and (c) demonstrate. In Fig.\ref{fig:three}(b) the cut from the upper part of the map is shown where it is seen that the other split component of $d_{xz,yz}$-dispersion is strong and thus complicates the LEG analysis. In Fig.\ref{fig:three}(c), in contrast, the intensity from these states is weak and the dispersing features responsible for the gapped FS are more pronounced. The presence of the gap is evident from Fig. \ref{fig:three}(e) which shows k$_F$-EDCs from intensity distribution along the line going trough the $\Gamma$-point (Fig. \ref{fig:three}(d)) measured above and below T$_c$. Leading edge shift between these EDCs is 0.8 meV. In order to estimate the SC gap anisotropy on this FS, we have extracted binding energy of the leading edge from the exemplary EDCs corresponding to the red markers in Fig.\ref{fig:three}(a). The result is shown in Fig. \ref{fig:three}(f). Also from this figure one clearly notices that gap on the hole-like pocket is anisotropic, with a maximum located on the shorter ellipse axis and minimum on the longer one. Fitting the data with a periodic function yields a difference between extrema of 0.75 meV.

Near the \(\Gamma\)-point hole-like pocket becomes too small (about 0.06 ${\text{Å}}^{-1}$ in diameter) to disentangle two components of the Fermi surface originated from two domain orientations. Analysis of the asymmetry of superconducting gap becomes very complicated and model-dependent. The presence of the gap itself is though apparent. This follows also from the typical leading edge position behavior extracted from the cut measured through the pocket center Fig. \ref{fig:three}(h). Presence of a deep minimum in this curve points to the back-folding of the dispersion due to superconductivity. If the top of the dispersion is located close to the Fermi level, as is the case here, opening of the gap results exactly in this behavior of the leading edge position \cite{evtushinsky2011fusion}. In the case of absence of the gap, i.e. nodes, one would expect a flat shape of this curve dictated mostly by the Fermi function since the spectral function is nearly equally strong also in between the Fermi level crossings due to proximity of the top.

Fig.4 summarizes our findings as regards the 3D gap function in FeSe. In this figure we show only the FS corresponding to single domain. The superconducting gap on both parts of the Fermi surface is anisotropic in k$_x$-k$_y$ plane as well as in k$_z$ direction. The largest leading edge gap is on the hole-like pocket near Z-point. The gap oscillations on this pocket have C$_2$ symmetry with maximum on a short axis of the ellipse. Near \(\Gamma\)-point the gap seems to be considerably smaller but not zero. A symmetry of the leading edge gap on the electron-like pocket also has C$_2$ behavior on each of the two ellipses. Its highest value corresponds to short axis of the ellipse and is smaller than one on the hole-like pocket. The behavior of the gap on the electron-like pocket is qualitatively the same for all k$_z$, but is characterized by a non-monotonic amplitude of the oscillations when going from A to M-point.

\begin{figure}[t]
	\centering
    \includegraphics[width=1\linewidth]{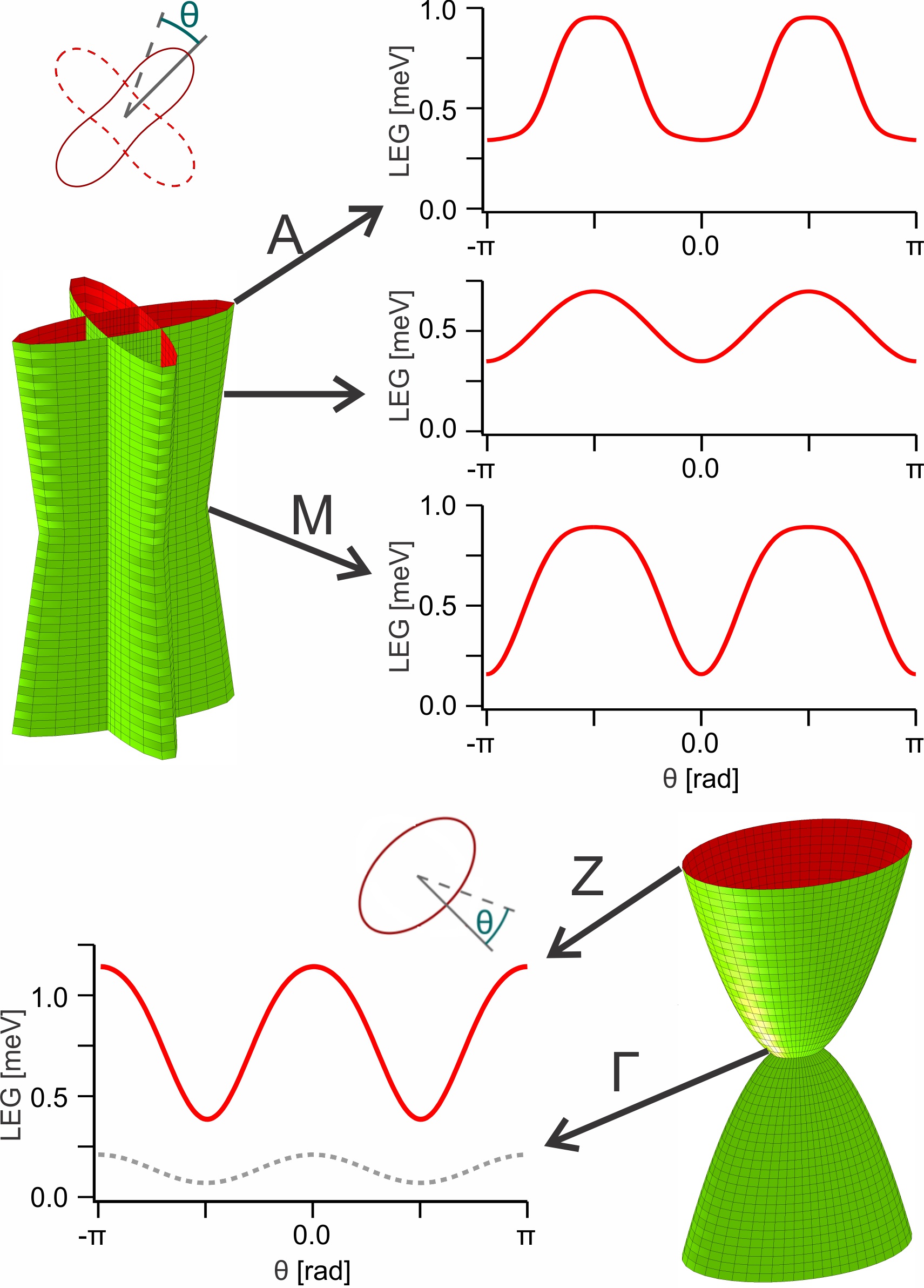}
	\caption{ Summary of the obtained results. The fits to the experimental LEG distribution in 3D BZ are shown for different k$_z$ values for hole and electron pockets. In both cases zero $\theta$ corresponds to the diagonal of the BZ, i.e. the direction in the momentum space which connects the center of hole-like pocket with the center of electron-like pocket.}
	\label{fig:four}
\end{figure}

We have previously detected a correlation between the size of the gap and degree of spin-orbit splitting in all main representatives of the iron-based superconductors \cite{Borisenko16NPh}. The present study confirms this with a new level of precision for FeSe. Indeed, the absolute value of the gap in the center of the BZ, where the spin-orbit splitting is maximal, is larger than the one on the electron-like pockets. Another correlation with the absolute value of the gap has been noticed by us in hole-doped 122 materials \cite{RN110}. There we have demonstrated that the gap is always the largest for the $d_{xz,yz}$-states and decreases as soon as the other orbital character is admixed.

We also compare the earlier determined anisotropic gap function in LiFeAs \cite{BorisenkoSymm} with the one determined in the present study in FeSe. Apparently, very different electronic structures result in qualitatively different gap structures. In the case of LiFeAs electron-like pockets are significantly larger and the gap oscillates on them in phase, having C$_4$ symmetry, i.e. it is maximal on both of them when crossing diagonal of the BZ and is minimal in between, regardless of shorter or longer axes of the ellipses. The gap of $d_{xz,yz}$ states is also the largest in FeSe but oscillates in the same way as the gap on the large $d_{xy}$-pocket in LiFeAs (which is absent in FeSe) having minima on the diagonals of the BZ. Oscillations of the gap on small and 3D $d_{xz,yz}$-pocket in LiFeAs have not been resolved. This comparison calls for detailed and quantitative theoretical estimates of the gaps in FeSe by the same methods applied to LiFeAs earlier \cite{RN112,RN105,RN109}. 

Presented results imply a significant anisotropy of the superconducting gap in FeSe, not only in-plane, but also as a function of k$_z$. This is in contrast to the expectations of the conventional spin-fluctuations mediated pairing theories where mostly isotropic s-wave gaps are expected, or anisotropy is different. The concept of orbital-selective Cooper pairing suggested recently \cite{sprau2017discovery} seems to be in agreement with our observations. There the concentration of the pairing in the particular orbital channel may arise from differences in the correlation strength for electrons with different orbital character. In particular, it explains persistently smaller $d_{xy}$-gaps observed experimentally by increased incoherence which would suppress the pairing within an itinerant picture. Indeed, the anisotropy of the superconducting gap in FeSe studied by us, can roughly be explained by the orbital composition of the states forming the Fermi surface in the normal state. As soon as the contribution of $d_{xz,yz}$ character is stronger - the gap reaches its maximum. On the other hand, this concept also requires adding a phenomenologically different quasiparticle spectral weights for the $d_{xz}$ and $d_{yz}$ orbitals. We do not observe significantly different Z-weights of these orbitals, because both electron-like pockets are present within a single domain and the corresponding peaks of the spectral function are equally sharp \footnotemark[100]. At the same time, our results are in a qualitative agreement with the gap anisotropy extracted from the tunneling data \cite{sprau2017discovery} with the difference in the absolute values being probably due to mentioned peculiarities of the LEG.

Our data are also in agreement with the variations of the pairing gap caused by nematicity itself \cite{Kang_arxiv}. In this approach the anisotropy of the gap arises from the mixing of $s$-wave and $d$-wave pairing channels without the necessity to postulate different Z-factors for each orbital.

In order to make a more rigorous statement as for the application of one or another theoretical approach, more detailed calculations are obviously needed to reproduce the whole 3D momentum dependence of the superconducting gap in FeSe determined in the present study.

This work was supported by Deutsche Forschungsgemeinschaft Grants No. BO1912/6-1 and No. BO1912/7-1. We acknowledge Diamond Light Source for time on Beamline I05 under proposals SI11643-1 and SI18586-1. We are grateful  to Dirk Morr, Matthew Watson, Andrey Chubukov and Alexander Kordyuk for the fruitful discussions. In addition, S. Aswartham and I. Morozov express their gratitude to the Volkswagen Foundation for financial support and are grateful to Dmitriy Chareev for valuable consultations on growing FeSe single crystals.

\footnotetext[100]{Supplemental Material}

\footnotetext[101]{It is known that Z- and \(\Gamma\)-point in FeSe correspond to photon energy 23 eV and 37 eV respectively \cite{maletz14PRB,watson2015suppression,watson15PRB}. However, when one probes electrons emitted away from the normal to the surface, the out-of-plane component (k$_z$) is not constant and decreases. In order to compensate for this when probing the part of the BZ near A- and M-points, one should use photons with different energies: 28 eV for A-point and 42 eV for M-point \cite{fedorov16}}

\bibliography{literature}
\newpage

{\bf 3D superconducting gap in FeSe from ARPES (Supplemental Material)}

Here we present ARPES data which demonstrate that electron-like Fermi surface in nematic state actually consists of 4 peanuts: one pair of orthogonal peanuts for one domain orientation and another pair for another domain.

\begin{figure*}[]
	\centering
    \includegraphics[width=0.9\linewidth]{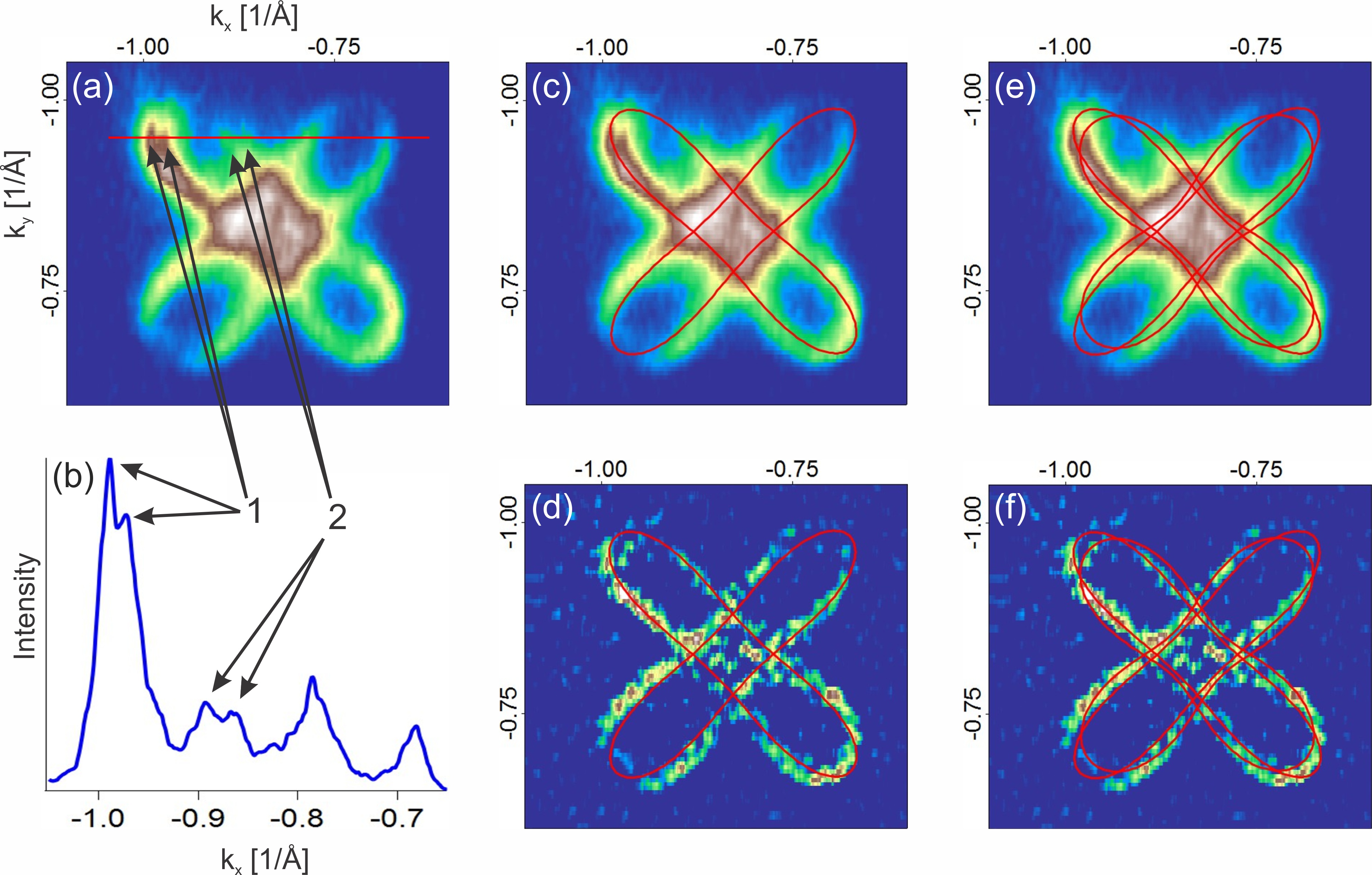}
	\caption{ (a) Fermi surface map of the electron-like pocket near A-point. (c),(e) one pair and two pairs of peanut-like shapes fitted to the pocket shape, respectively. (d),(f) the same curves as in (c),(e) but on processed images of the map. These images were obtained by taking the 2-nd derivative, which emphasizes weak features.}
	\label{fig:S_one}
\end{figure*}

At a first glance, the electron-like FS in Fig.\ref{fig:S_one}(a) and in the maps shown in the main text indeed consists of two peanuts. Upon closer visual inspection the map reveals more complex structure. The doubling is clearly seen in the regions of the k-space shown by black arrows. These double-features are even better seen as peaks in momentum distribution curve taken along the red line on the map Fig.\ref{fig:S_one}(b). Another observation is the shape of the pockets themselves. They appear to be different.  If we try to fit a pair of identical peanut-like contours to the pocket shape, it is not possible to obtain a good fit for all pockets simultaneously. For instance, one can fit quite well the upper-left petal (Fig. \ref{fig:S_one}(c),(d)), but lower-left and lower-right ones turn out to be too wide. Both mentioned arguments for doubling are also confirmed by carefull inspection of the maps presented in the main text. From this we can conclude that the measured map is a superposition of ellipses with different width, with some parts of them being suppressed by matrix element effects. Indeed, if we add a second pair of peanut-like contours, we can fit them to all experimental features on the map, including both doubled features in the upper-left petal (Fig. \ref{fig:S_one}(e),(f)). This result is in an agreement with DFT calculations for the orthorhombic state (see Fig.2 (b-d) in Ref. \onlinecite{fedorov16}) as well as with the earlier experimental data (see Fig.2 (e,f) in Ref. \onlinecite{fedorov16}). We would also like to note, that the ellipses do not actually cross each other because of spin-orbit interaction. Since such anticrossing gaps are very small, they are not directly resolved in the maps. Strictly speaking, there is one small and one big electron pocket (and another pair of those if we are in a nematic state) which only resemble crossed peanuts.

\end{document}